\documentclass[final,5p,times,twocolumn]{elsarticle}

\usepackage{amssymb}
\usepackage{amsmath}

\usepackage{float} 

\usepackage{hyperref}
\hypersetup{
    colorlinks=true,
    linkcolor=black,
    citecolor=black,
    filecolor=black,
    urlcolor=black,
}

\usepackage{color,soul}

\biboptions{numbers,sort&compress}

\journal{Elsevier}

\begin{document}

\begin{frontmatter}


\title{Prestrain-induced bandgap tuning in 3D-printed tensegrity-inspired lattice structures}

\author[caltech]{Kirsti Pajunen\corref{cor1}}
\ead{kpajunen@alumni.caltech.edu}
\author[sbu]{Paolo Celli}
\author[caltech]{Chiara Daraio}

\address[caltech]{Division of Engineering and Applied Science, California Institute of Technology, Pasadena, CA 91125, USA}

\address[sbu]{Department of Civil Engineering, Stony Brook University, Stony Brook, NY 11794, USA}

\cortext[cor1]{Corresponding author}

\begin{abstract}

In this letter, we provide experimental evidence of bandgap tunability with global prestrain in additively-manufactured tensegrity-inspired lattice structures. These lattices are extremely lightweight and designed to exhibit a nonlinear compressive response that mimics that of a tensegrity structure. We fabricate them out of a stiff polymer but, owing to their peculiar design, they are compliant and remain elastic up to high levels of precompression. In turn, unlike tunable metamaterials made of soft polymers, the response of our lattices is not dominated by damping. We perform experiments on a one-dimensional lattice subject to longitudinal wave excitation and varying levels of static longitudinal precompression, and observe continuous tuning of both the wave speed and the location and width of the lowest-frequency bandgap. 

\vspace{15px}
\noindent{\textbf{This article may be downloaded for personal use only. Any other use requires prior permission of the authors and Elsevier Publishing. This article appeared in}: \emph{Extreme Mechanics Letters} 44, 101236 (2021) \textbf{and may be found at}: \url{https://doi.org/10.1016/j.eml.2021.101236}}
\vspace{10px}

\end{abstract}


\end{frontmatter}


\section{Introduction}

Lattice structures are lightweight foam-like mechanical systems featuring architected assemblies of simple structural elements~\cite{Fleck2010}. Known to most for their static properties, they also present interesting dynamic attributes and elastic wave dispersion characteristics. For example, they can be engineered to propagate elastic waves at desired direction-dependent wave speeds~\cite{Phani2006, Gonella2008, Casadei2013, Matlack2016, Bayat2017, Bacigalupo2017}, and they can be designed to feature bandgaps, i.e., frequency ranges of strong wave attenuation~\cite{Phani2006, Gonella2008, Liebold2014, Kroedel2014, Wang2015}. In classical periodic lattice structures, bandgaps are predominantly of the Bragg scattering type~\cite{Phani2006}; when lattices feature additional/auxiliary structural elements located within each unit cell, they also feature locally-resonant bandgaps at the resonance frequencies of said auxiliary microstructures~\cite{Gonella2009, Kroedel2014, Matlack2016, Tallarico2017}. Typically, the wave properties of lattices are set in stone after fabrication. This limitation can be lifted for specific lattice structures made of soft materials that display buckling-induced phase transitions and can therefore change shape in response to external precompression~\cite{Bertoldi2008, Wang2012, Wang2014, Shan2014, Pal2016, Li2018}. These lattices typically feature a discrete degree of tunability, since they can only transform into a finite number of geometrical configurations when loaded. Moreover, their soft nature makes practical realizations challenging~\cite{Shan2014}, since damping tends to dominate these systems' wave response~\cite{Raney2016}. Continuous tunability, whereby the wave properties can be swept within a geometry-dependent interval, is usually not attainable in lattice structures, unless they are soft~\cite{Chen2017}, or endowed with multifunctional capabilities. In this case, tunability can be induced by an external stimulus such as an electric field~\cite{Nouh2015, Celli2017}, temperature~\cite{Zhang2016, Nimmagadda2019} or a magnetic field~\cite{Bayat2015, Wang2016}.

Tensegrity lattices have recently emerged as systems that are made of stiff materials and yet are characterized by continuous wave tunability in response to prestrains~{\cite{Skelton2001, Krushynska2018, Amendola2018, Pal2018, Liu2019, fraternali2012solitary, fraternali2014multiscale, davini2016impulsive, rimoli2017mechanical, rimoli2018reduced, Wang2018wave}}. These lattices are made by the repetition of a tensegrity unit cell featuring pin-jointed and prestressed arrangements of cables and rigid struts~\cite{Skelton2001, zhang2015tensegrity, masic2006optimization, oppenheim1997mechanics, tibert2003review}. From a statics perspective, they are lightweight~\cite{nagase2014minimal, Skelton2001, zhang2015tensegrity}, capable of supporting large global prestrains~\cite{amendola2014experimental} and exhibit load-limiting, non-linear characteristics~\cite{oppenheim2000geometric, Skelton2001, rimoli2017mechanical, mascolo2018geometrically, amendola2014experimental, fraternali2015mechanical}. This extreme deformability is at the basis of their wave tunability attributes: their dispersion characteristics can be significantly altered by varying the degree of local~\cite{Krushynska2018, Amendola2018, Pal2018, Liu2019} or global prestrain~{\cite{Amendola2018, Wang2018wave}}. To date, an experimental demonstration of this wave tunability is lacking.

In this work, we study the wave response of 3D-printable, architected lattices that are designed to mimic several appealing characteristics of tensegrities: they are lightweight, exhibit nonlinear load limitation from local member buckling, and remain elastic and intact under large global compressive prestrains. These lattices are printed with a single stiff polymer (polyamide) and their architecture is tailored to exhibit a tensegrity-analog static response~\cite{Pajunen2019}. We tile 3D unit cells into one-dimensional arrays and test their longitudinal wave mechanics. Leveraging these characteristics and the low levels of damping intrinsic of stiff materials, we provide an experimental demonstration of elastic wave tuning in response to global prestrains. In particular, we show that both the wave speed and the width of the Bragg scattering bandgap can be continuously tuned via precompression.

\section{Basic building blocks and quasistatic response}

The 3D-printed tensegrity-inspired structure we use in this work is introduced in detail in Ref.~\cite{Pajunen2019}. It has 6 square, orthogonal faces for tessellation in three dimensions. Its geometry is analagous to a truncated octahedron tensegrity unit cell, which is introduced and studied in Ref.~\cite{rimoli2017mechanical}. Although the 3D-printed structure has fixed joints, it has an equivalent quasistatic and impact response to a pin-jointed, buckling tensegrity unit cell. These structures are very lightweight, with relative density of 2.5\%.

\begin{figure}[!htb]
\centering
\includegraphics[width=\columnwidth]{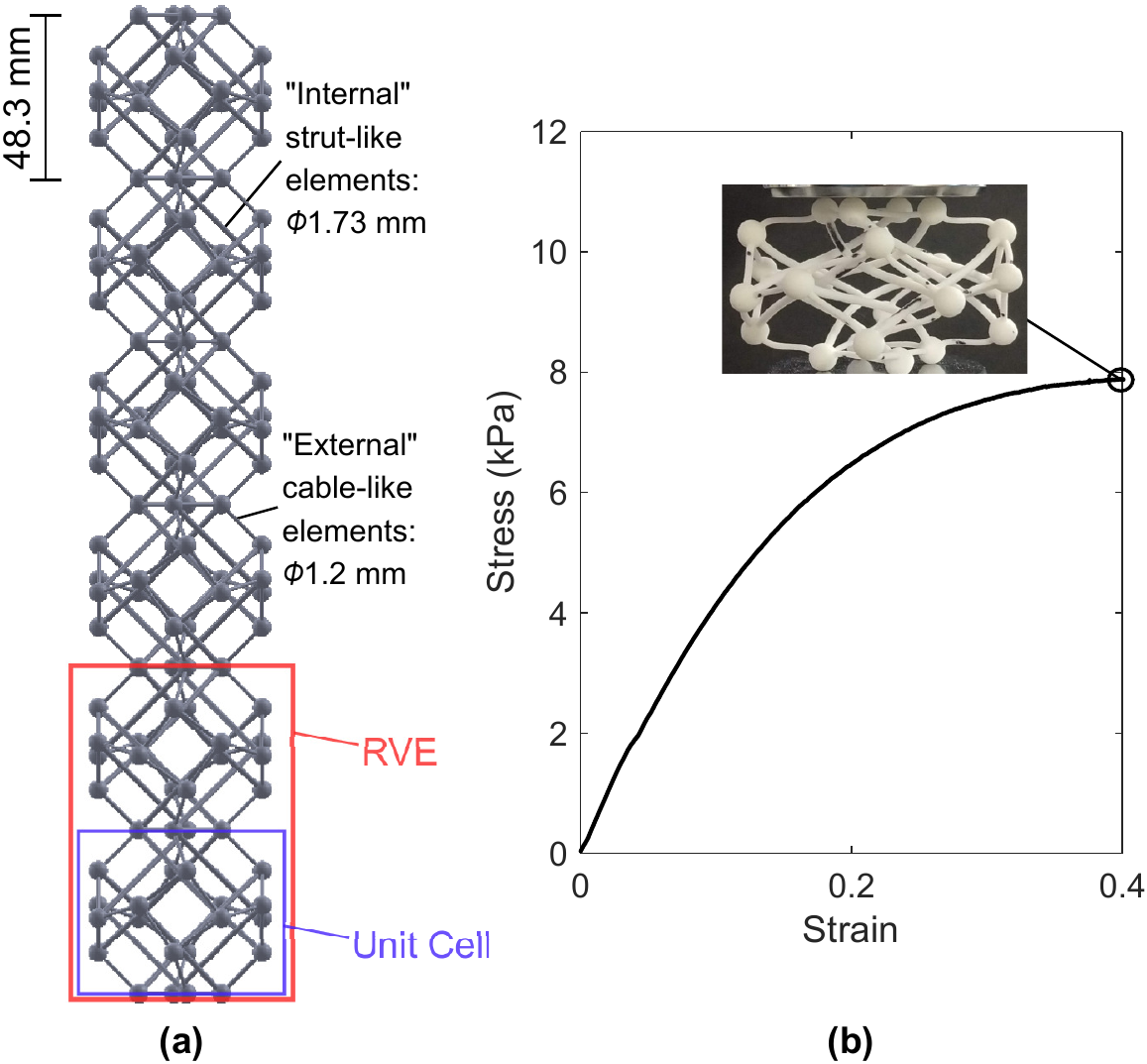}
\caption{(a) A 3-RVE 1D lattice with dimensions. The 1D representative volume element, or RVE, is enclosed in the red box, {and its building block unit cell is enclosed in the blue box.} (b) The experimental, global stress-strain response of a tensegrity-inspired unit cell, with its deformation shown at 0.4 strain.}
\label{FigureQS}
\end{figure}

A 1D lattice comprised of six tensegrity-inspired units is shown in Figure~\ref{FigureQS}(a). 
In each unit, the thin cable-like {elements} are continuously connected at spherical joints on the outer ``surface'' of the structure, while the bulkier strut-like {elements} run within the structure and are connected via {the cable-like elements}. {Similarly, a continuous web of cables connecting discontinuous struts} is a characteristic of tensegrity systems~\cite{zhang2015tensegrity}. The dimensions of the fabricated structures are the following: the distance between flat square faces (the size of a unit) is $L=48.3\,\mathrm{mm}$, the strut{-like element} diameter is $d_s=1.7\,\mathrm{mm}$, the cable{-like element} diameter is $d_c=1.2\,\mathrm{mm}$, and the diameter of a spherical joint is $d_j=5.7\,\mathrm{mm}$.
To tessellate a unit into a one-dimensional lattice, we need to first construct a one-dimensional representative volume element (RVE). The faces of a unit are twisted with respect to their perpendicular axes. Because of this, the unit is not directly tessellatable without a series of reflection operations, as described in Ref.~\cite{rimoli2017mechanical}. The one-dimensional RVE, therefore, is obtained by merging a unit and its reflection with respect to the plane of the top face. The 1D RVE is enclosed in the red box in Figure~\ref{FigureQS}(a).

Our structures are fabricated from the Shapeways.com\textsuperscript{\textcopyright} polyamide PA2200 material using selective laser sintering. First, we perform a quasi-static compression experiment on one RVE. Owing to the fact that the struts undergo compression and subsequent buckling, the structure exhibits a plateau in the global stress with increasing compressive strain, as shown in Figure~\ref{FigureQS}(b). Stress is calculated from the reaction force exhibited by the bottom four nodes, divided by the square projected area of the cell. Strain is calculated as the change in the structure height divided by its original value. Quasistatic tests are performed on an Instron E3000 testing machine. We can observe that, like classical tensegrity systems, our printed structure also remains elastic under severe deformation, exhibits a nonlinear response, and has a very low relative density. Also, it exhibits post-buckling stability and load limitation.


\section{Methods}

\subsection{Experimental setup}

\begin{figure*}[!htb]
\centering
\includegraphics[width=\textwidth]{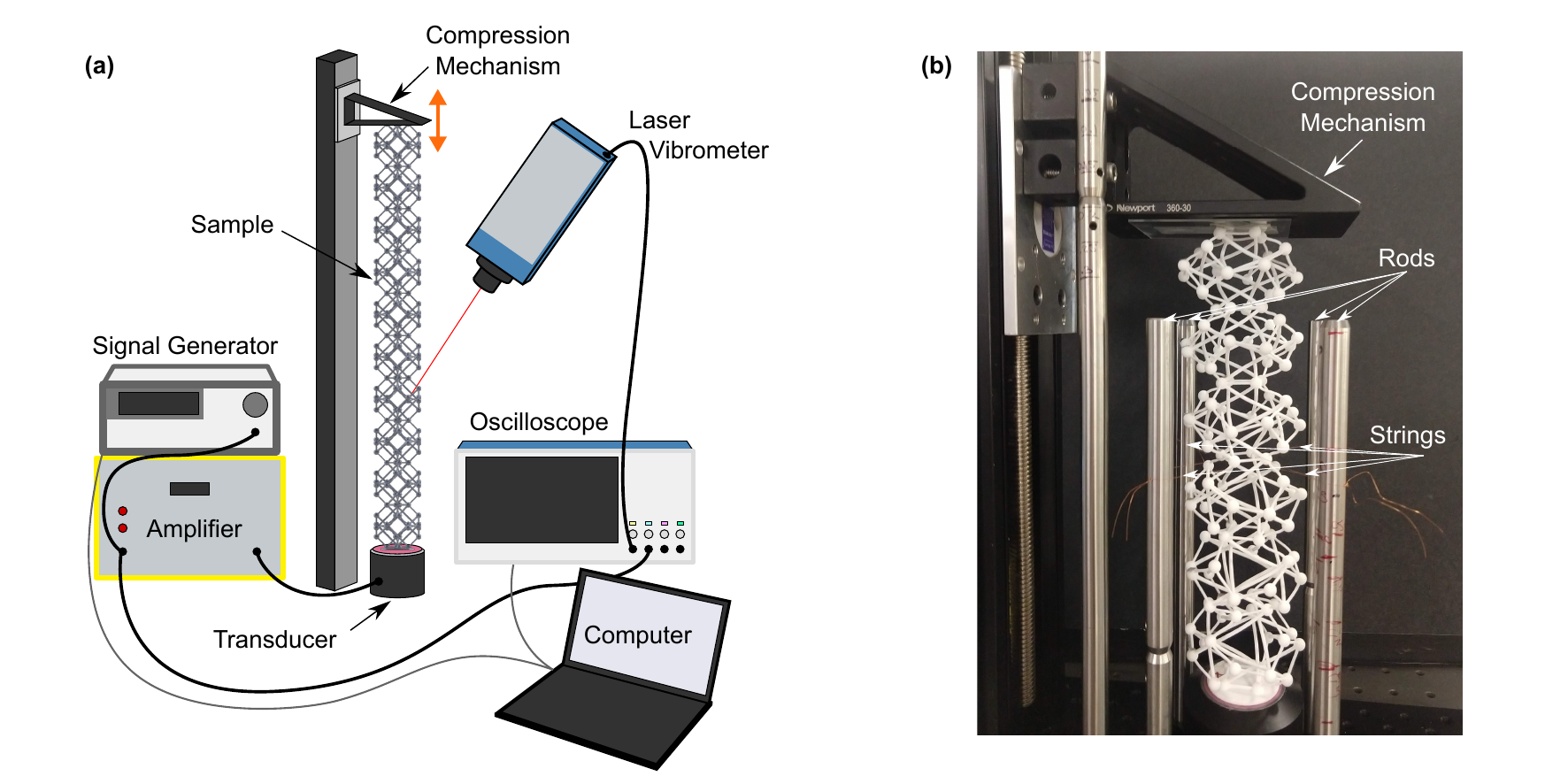}
\caption{(a) Schematic of the experimental setup used to study the spectro-spatial wave characteristics of a tensegrity-inspired 1D lattice. (b) Photo of the apparatus designed to prevent global buckling during compression of the lattice. Strings are used to hold the lattice in-axis while still allowing vertical displacement during longitudinal wave propagation. A compression level of 20\% strain is shown.}
\label{FigureSetup}
\end{figure*}

To study the dynamic frequency response of the tensegrity-inspired structure, we resort to the experimental setup shown in Figure~\ref{FigureSetup}(a). 
Throughout this study, we use two one-dimensional lattice specimens, with 5 and 3 RVEs. We intend to measure both their transmission characteristics (to identify bandgaps) and their dispersion properties. The latter task requires us to measure the response at multiple locations along the direction of wave propagation at each RVE. We fabricate the lattice with a plate on the bottom face that intersects the four spherical nodes to simplify its attachment to the wave source. 

We devise a compression apparatus to study the effect of global prestrain on the dynamic response of the structure. A 90-degree metal bracket mounted on a linear stage (Velmex MN10) can move continuously in the vertical direction. The flat surface of the bracket compresses the top face of the structure, controlling the applied strain. Our samples are long and slender, and global lateral buckling can occur during their compression. To prevent this undesired behavior, and to compress the structure to high strains, we only compress 3-RVE-long specimens and we construct an apparatus that holds the structure in its vertical axis, as shown in Figure~\ref{FigureSetup}(b). The apparatus uses four tensioned strings, each with one extreme tied to one of four equidistant locations on the lattice, and the other extreme tied to, but allowed to slide along, a vertical rod. The strings hold the structure in place only laterally and have minimal influence on longitudinal wave propagation.

To propagate longitudinal waves along the structure, we connect a piezoelectric transducer (Panametrics V1011) to the bottom plate on the sample. We excite it with a one-cycle wide-band burst with a carrier frequency of 200 Hz. Using a single-point Laser Doppler Vibrometer (LDV, Polytec CLV-2534), we measure the velocity time history along the specimen's length at each intersection between tensegrity-like units, as well as on the bottom plate. This measurement is repeated 128 times at each location and averaged to improve the signal-to-noise ratio. We also employ a high-pass filter to eliminate all ambient noise at frequencies lower than 100 Hz. Once we collect the measured data into a time-space matrix, we obtain a frequency-wavenumber data matrix by using a 2D Discrete Fourier Transform (2D-DFT). We zero-pad the data prior to performing the 2D-DFT operation, to interpolate along the wavenumber direction and improve results visualization despite having only few spatial samples. We also obtain frequency transmission data and space-time diagrams through post-processing of the output velocity data. These experimental and post-processing methods have been similarly implemented in Ref.~\cite{Palermo2019}.

\subsection{Numerical modeling}
\label{s:num}

{To provide a comparison with the experimental results as well as further insight into the wave response of our structure, we conduct finite element simulations in COMSOL}. We construct a 3D model of a 1D RVE and mesh it with tetrahedral volume elements. The material properties are obtained from tensile tests on ASTM D638 material test specimens using an Instron E3000. The material exhibits a hyperelastic response, straying from a linear elastic response at around 4\% strain. {By extracting the linear portion of the experimental response, we select a linear elastic material model with Young's Modulus of 1.29~GPa, Poisson's ratio of 0.3, and density of 930~kg/m\textsuperscript{3}}. To calculate the dispersion relation of the 1D lattice, we use an eigenfrequency step with Bloch periodic boundary conditions, thus simulating the response of an infinite lattice. To create a tessellatable unit, we make a centered cut through the top and bottom faces to create flat surfaces on which to apply the vertical periodic boundary conditions.

To produce the dispersion curves, the software solves an eigenvalue problem for each wavenumber value $k$ in the irreducible Brillouin zone (IBZ, 0--$1/2a$ $\mathrm{1/m}$), where $a=2L$ is the length of the 1D RVE. To calculate the dispersion curves for varying levels of precompression, we first use a stationary step to solve for the quasistatic compressive response. Then, the final conditions from the stationary step (including the stresses arising from precompression) are set as the initial conditions for the linear eigenfrequency step. This small-on-large approach is standard procedure when modeling bandgap tunability due to mechanical forces~\cite{Bertoldi2008, Li2018}.

Finally, a finite lattice with 3 RVEs is simulated in COMSOL in order to find the longitudinal frequency transmission response. Since this simulation takes into account the finite size and boundaries of the experiment, we expect it to capture the experimental response better than the numerical dispersion curves. 
In particular, we perform a harmonic analysis (from 0 to 500 Hz) by applying a base excitation with amplitude of 1~mm to the bottom face surface, while the top face surface is kept fixed as in the experiment. The input and output vertical displacement amplitudes are extracted at approximately the same locations as in the experiment. In the simulations with precompression, the nonlinear quasistatic step is performed before the harmonic analysis.

\section{Results}

\subsection{Unstrained lattice response}

\begin{figure*}[!htb]
\centering
\includegraphics[width=\textwidth]{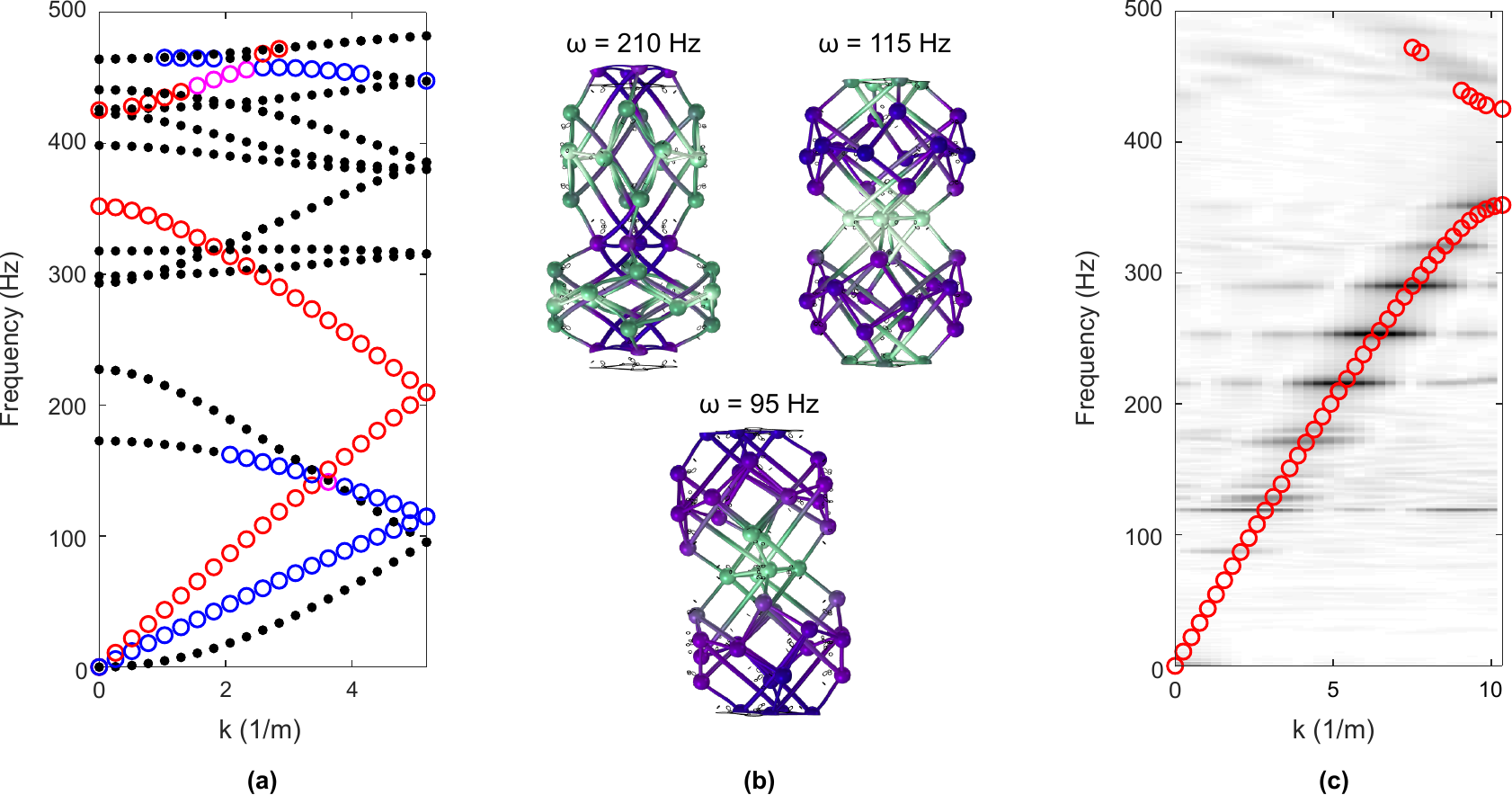}
\caption{Unstrained 1D lattice response. (a) Numerical dispersion relation. The red circular markers are longitudinal modes, the magenta ones are combined longitudinal and rotational modes, the blue ones are rotational modes, and the black dots indicate modes that do not clearly belong to any of the above categories, such as flexural modes. (b) Examples of longitudinal (top left), rotational (top right), and flexural (bottom) mode shapes at the edge of the IBZ. (c) Experimentally-reconstructed dispersion plot for longitudinal wave motion, for the 5-RVE lattice. We expect the dispersion branches to follow the maxima of the colormap~\cite{Celli2019}. The red circles indicate the ``unwrapped'' longitudinal wave modes from the numerics.}
\label{FigureUnstrained}
\end{figure*}

The characteristic dispersion curve obtained by simulating the infinite lattice is shown in Figure \ref{FigureUnstrained}(a). The markers are color-coded according to the mode of wave propagation they correspond to. We consider the mode shape associated with each single eigenvalue, and we extract the vertical displacement amplitudes at the top and middle of the RVE and the curl amplitude around the vertical axis. We establish quantitative thresholds based on the ratios of these values and color the corresponding eigenvalue marker accordingly~\cite{Kroedel2018}. Examples of the mode shapes for longitudinal, rotational, and flexural modes at the edge of the IBZ are shown in Figure \ref{FigureUnstrained}(b). From previous works we know that, as the structure is compressed, vertical deformation is coupled with rotation of the nodes~\cite{Pajunen2019}. This explains the coupling between modes that is observed above 420~Hz. Up until that frequency, however, the path followed by the longitudinal mode (red markers) is unambiguous. In particular, we see that a bandgap exists between 351 and 425~Hz. The fact that the low-frequency branch folds around the edge of the Brillouin zone, and the observation that the branches before and after the gap resemble the acoustic and optical modes of a diatomic system (in terms of shape and slope) suggest that this bandgap is primarily due to Bragg scattering effects~\cite{Brillouin1953}.

The experimentally reconstructed dispersion relation of the unstrained 5-RVE lattice is given by the grayscale colormap of Figure~\ref{FigureUnstrained}(c). The dispersion branches are expected to connect the locations of high amplitude, which correspond to structural resonances of the finite specimen~\cite{Celli2019}. Since we have two measurement points per unit cell, one every $a/2$, the plot extends to $k=1/a$ instead of $1/(2a)$. Thus, prior to overlapping the longitudinal numerical curve onto the experimental data, we ``unwrap'' it about $1/(2a)$~\cite{Dontsov2013}, obtaining the red markers of Figure~\ref{FigureUnstrained}(c). The experimental and numerical results agree, with the numerical dispersion following the maxima of the colormap. In particular, the slopes of the lower branch coincide between the two sets of results. The bandgap, highlighted in the experimental plot by the absence of dark regions for vast frequency ranges, fall in a similar range of frequencies. Note that the region below 100~Hz is affected by high-pass filtering. {We identify two sources of differences between numerics and experiments. First, the experimental results are for a finite lattice, and thus are subject to boundary effects, whereas the numerical results are for an infinite lattice. Second, since the laser vibrometer is not perfectly parallel to the length of the chain, small lateral deformations may also be detected in the experimental results. Even with these factors, the simulation and experimental results agree well.}

\subsection{Influence of global prestrain}

\begin{figure*}[!htb]
\centering
\includegraphics[width=\textwidth]{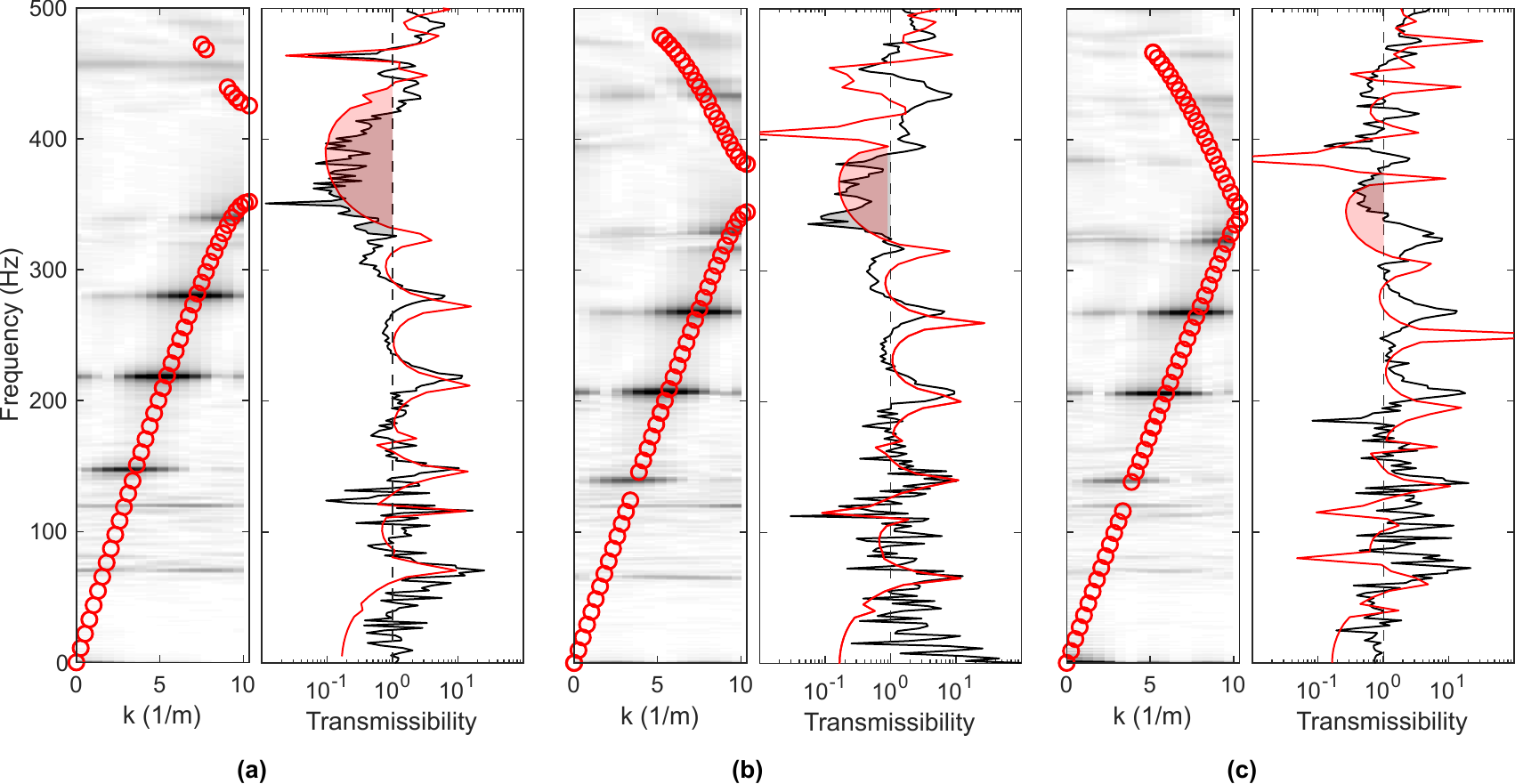}
\caption{Precompression-induced tunability. The left panel of each subfigure represents the experimentally-reconstructed dispersion curve (gray colormap), with overlapped circular markers corresponding to the numerical dispersion relation. The right panel is the transmissibility plot, where the experimental curves are black and the numerical ones are red. The dashed vertical line indicates a transmissibility of 1. The shaded gray and red regions serve as a guide to the eye and indicate what we identify as bandgaps for the experimental and numerical curves, respectively. (a) Lattice compressed to 0\% global strain. (b) 5\% strain.  (c) 10\% strain.}
\label{FigurePrestrain}
\end{figure*}

Since the baseline structure exhibits a nonlinear load-limiting response, we expect the dispersion characteristics of the 1D lattice to change with increasing axial compression. To examine this, we perform experiments on a 3-RVE lattice at 0\%, 5\%, 10\%, 15\%, and 20\% global compressive prestrains. Above 20\%, the structure globally deforms into its second buckling mode. Note that this could have been prevented by adding additional supporting strings to our setup. The transmissibility is calculated as the ratio of the velocity at the final measurement point (output) to the velocity at the first measurement point (input). The experimental and numerical dispersion curves and transmissibility plots are shown in Figure~\ref{FigurePrestrain} for 0\%, 5\%, and 10\% strain. Numerical results are shown in red, and the experimental results in black.

Looking at the numerical dispersion curves we clearly see that, with increased compressive strain, the bandgap width reduces and the slope of the acoustic branch decreases, lowering the onset of the bandgap. The only signature of bandgap closure that can be read from the experimentally-reconstructed dispersion plots is the narrowing of the peak-less frequency region, that extended from 340 to 440~Hz in Figure~\ref{FigurePrestrain}(a) for 0\% strain. A perfect match is not attainable here because the numerics represent the response of an infinite lattice. 

The transmissibility plots, being representative of a finite-lattice response, show a better match between numerics and experiments. We first analyze the 0\% strain case of Figure~\ref{FigurePrestrain}(a), where we can see that the structural resonance peaks between black and red curves almost coincide (with a minimal shift of 8~Hz, and with the experimental response appearing to be stiffer). The noisy nature of the experimental response below 100~Hz is again due to high-pass filtering. A strong anti-resonance is visible for both numerics and experiments at around 450~Hz. From these curves, we identify as a bandgap the region between peaks at approximately 340 and 450 Hz, where the transmission dips below $10^0$. This attenuation region, shaded in red for the numerics and in gray for the experiments, is in the vicinity of the numerical dispersion bandgap, albeit slightly wider. As we increase the strain to 5\% and 10\%, as shown in Figures~~\ref{FigurePrestrain}(b) and (c), respectively, the peaks associated with the modes of the finite structure shift towards lower frequencies. The numerical curves show an increased peak density near the bandgap: the valley identified as gap in Figure~\ref{FigurePrestrain}(a) becomes increasingly narrow and shallow (see the evolution of the shaded regions), confirming that prestrain causes the bandgap to diminish in size. We can also note that, as we increase prestrain, the shift between numerical and experimental resonances increases (reaching $19~\mathrm{Hz}$ in Figure~\ref{FigurePrestrain}(c)). This is not surprising, as the difference between numerics and experiments is bound to increase for larger prestrains, for several reasons. First, the material model used in COMSOL is linear elastic. At higher compression levels, local strains may exceed the region of negligible deviance from the linear elastic region. Second, buckling of the structure could affect the response even before any buckling is visually apparent. However, even though these differences exist, numerical and experimental curves are comparable. 

\begin{figure}[!htb]
\centering
\includegraphics[width=\columnwidth]{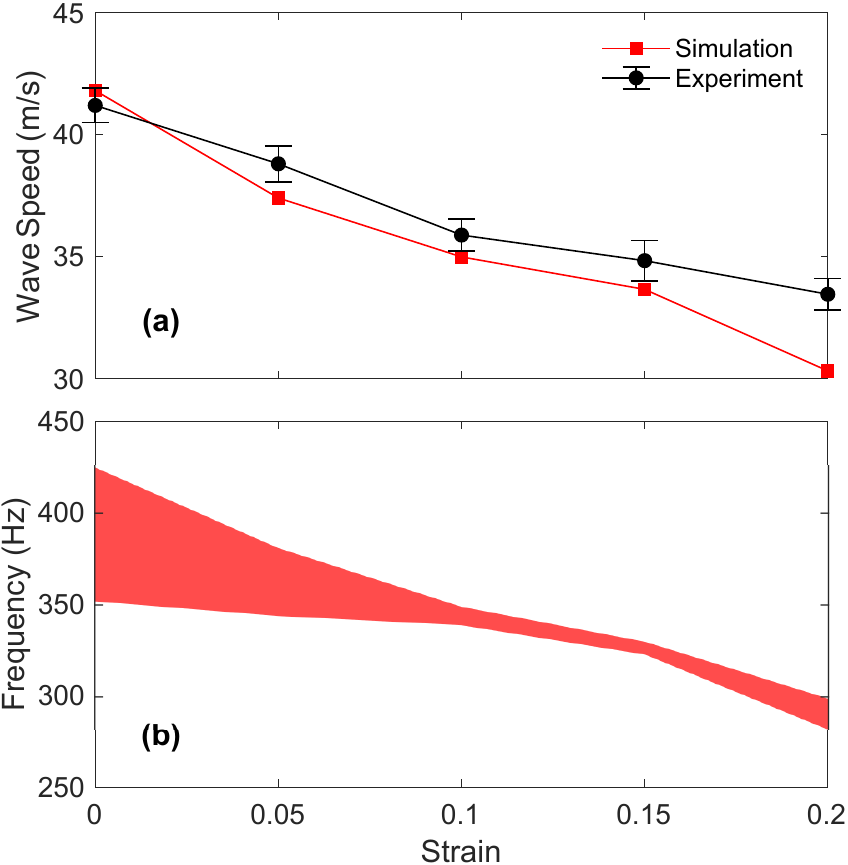}
\caption{(a) Experimental and numerical longitudinal wave speeds at varying levels of compression. The error bars on the experimental line indicate the wave speeds as calculated by the highest peak and lowest valley in the amplitude of the measured signal of the first transmitted wave. The black circle markers give the average of these two values. (b) Evolution of the bandgap width (shaded region) with strain, as predicted by the numerical dispersion curve.}
\label{FigureWS_BG}
\end{figure}

For higher levels of prestrain, namely 15\% and 20\%, identifying the bandgap is increasingly difficult. For this reason, we verify the consistency of numerics and experiments by tracking the evolution of the longitudinal wave speed with prestrain, as shown in Figure~\ref{FigureWS_BG}(a). The wave speed is extracted from the numerical results as the slope of the first mode of longitudinal wave propagation in the long-wavelength regime (small $k$). The experimental (phase) velocity is measured from the time histories recorded at the first and last measurement locations: we divide the spatial distance between those locations by the difference between time of occurrence of the same feature of the wave packet (a peak or a valley). The black circle marker is the average of the speed values obtained for the first peak and the first valley in the recorded signal. Since the packet is centered at 200~Hz, and since Figure~\ref{FigurePrestrain} shows that dispersive features appear around 300~Hz, we expect this speed to be characteristic of the non-dispersive part of the branch. 

From Figure~\ref{FigureWS_BG}(a), we see that the numerical and experimental wave speeds match well at 0\% strain. Their wave speeds reduce similarly from 0\% strain to 10\% strain, with the numerics showing a 16\% and experiments showing a 13\% reduction in wave speed. As compression is increased, the experimental results again show a stiffer response with respect to the numerics, resulting in higher wave speeds. From 0\% to 20\% strain, the numerical wave speed reduces by 27\%, and the experimental wave speed reduces by 19\%. Despite the quantitative discrepancies, our experimental results qualitatively confirm that wave speed and bandgap can be continuously tuned with prestrain in tensegrity-inspired lattice structures. 

Due to difficulties in bandgap identification for finite-size systems, we resort to numerics only to visualize in a single plot the bandgap evolution for higher prestrains. The evolution of the bandgap as derived from the numerical dispersion curve for our tensegrity-inspired lattice is shown in Figure \ref{FigureWS_BG}(b). At no precompression, the gap is 73 Hz wide, from 352 to 425 Hz. As precompression is applied, the bandgap narrows significantly and shifts to lower frequencies. Interestingly, at 15\% strain, the bandgap nearly closes and is only 7 Hz wide, from 323 to 330 Hz. The bandgap then slightly reopens at 20\% strain.

\section{Conclusions}

Recent numerical studies on tensegrity lattices have shown them to be tunable with compression and prestrain, while having the unique advantages of being highly elastically deformable and having extreme strength-to-weight attributes. In this work, we demonstrate the dispersion tunability attributes of tensegrity-inspired 3D-printable lattices. The nonlinearity of the compressive response of the lattice causes a dramatic evolution of its dynamic characteristics. Continuous tunability of the bandgap and wave speeds is obtained by increasing the level of global precompression. This tunability is achieved with a compliant structure made of a stiff material, whose response is not dominated by damping. Since the deformation remains elastic even at large strains, a repeatable tuning of the lattice response is achievable. While this preliminary study focuses on one-dimensional lattices, tunability can be extended to larger two- and three-dimensional assemblies of tensegrity-inspired cells, where the issues related to global buckling we encountered here are bound to disappear. This could open avenues for new metamaterials with tunable wave focusing and waveguiding attributes. Moreover, one could also envision using these tensegrity-inspired structures as springs connecting larger masses, {as illustrated in the work of Amendola et al.~\cite{Amendola2018} and Yin et al.~\cite{Yin2020}}, to create phononic systems with richer dynamics.

\section{Acknowledgements}
This research was conducted with Government support under and awarded by DoD, Air Force Office of Scientific Research, National Defense Science and Engineering Graduate (NDSEG) Fellowship, 32 CFR 168a.


\bibliographystyle{model1-num-names} 


\end{document}